\documentclass[twocolumn]{aa}
\usepackage{graphicx}
\usepackage{amssymb}
\usepackage{times}
\usepackage{natbib}
\bibpunct{(}{)}{;}{a}{}{,}
\usepackage{color}
\usepackage{xspace}
\usepackage{epsfig}
\usepackage{epstopdf}
\def\FB {F_\mathrm{B}}
\def\FS {F_\mathrm{S}}
\def\DS {D_\mathrm{S}}
\def\DL {D_\mathrm{L}}

\def\tE {t_\mathrm{E}}
\def\ThE {\theta_\mathrm{E}}
\def\RE {R_\mathrm{E}}

\def\ThS {\theta_\mathrm{*}}

\def\anger{\theta_\mathrm{E}}
\def\to {t_\mathrm{0}}
\def\uo {u_\mathrm{0}}
\def\MJ        {M_\mathrm{J}}
\def\mstar     {M}

\def\mpla      {m}
\def\apla      {a}

\def\msun      {M_{\odot}}
\def\mTer      {M_{\oplus}}
\def\rsun      {R_{\odot}}

\def\Epsdq     {{\varepsilon_\mathrm{m}(d,q)}}

\def\EpsphysM  {{\tilde{\varepsilon}_\mathrm{p}}(\apla,\mpla)}

\newcommand\Fig[1]  {Fig.~\ref{#1}}

\newcommand\Sec[1]  {Sec.~\ref{#1}}
\def\d                {\mathrm{d}} 
\newcommand\dix[1] {\times 10^{#1}}

\def\planet        {OGLE~2005--BLG--390Lb\xspace}
\def\idalin        {\citet{2005ApJ...626.1045I}\xspace}

\def\event         {OGLE~2005--BLG--390\xspace}
\def\system        {OGLE~2005-BLG-390L\xspace}
\def\source        {OGLE~2005-BLG-390\xspace}
\def\planetgould   {OGLE~2005-BLG-169Lb\xspace}
\def\idalin        {\citet{2005ApJ...626.1045I}\xspace}

\newcommand\dan[1] {{{\color{black} #1}}}

\begin{document}
\title{Limits on additional planetary  companions to \system}
\titlerunning{Multiplicity of  \system}
\authorrunning{D.~Kubas, A.~Cassan \emph{et al.} }
\author {D.~Kubas\inst{1},  A.~Cassan\inst{2}, M.~Dominik\inst{3}\thanks{Royal Society University Research Fellow},  
D.~P.~Bennett\inst{4}, J.~Wambsganss\inst{2}, S.~Brillant\inst{1}, J.P.~Beaulieu\inst{5},
M.D.~Albrow\inst{12},  V.~Batista\inst{5}, M.~Bode\inst{19}, D.M.~Bramich\inst{14}, 
M.~Burgdorf\inst{19},  J.A.R.~Caldwell\inst{18}, H. Calitz\inst{16}, K. H.~Cook\inst{21}, Ch.~Coutures\inst{8}, 
S.~Dieters\inst{9}, D.~Dominis Prester\inst{7}, J.~Donatowicz\inst{17},  P.~Fouqu\'e\inst{6}, J.~Greenhill\inst{9}, K.~Hill\inst{9},
M.~Hoffman\inst{16}, K.~Horne\inst{3}, U.G.~J{\o}rgensen\inst{13}, N.~Kains\inst{3},
S.~Kane\inst{20}, J.B.~Marquette\inst{4}, R.~Martin\inst{11}, P.~Meintjes\inst{16},  J.~Menzies\inst{10}, K.R.~Pollard\inst{11}, K.C.~Sahu\inst{15}, C.~Snodgrass\inst{1}, I.~Steele\inst{19}, Y.~Tsapras\inst{22}, C.~Vinter\inst{13} , A.~Williams\inst{11}, K.~Woller\inst{13} \and M.~Zub\inst{2}.\\
({\sc The PLANET/RoboNet Collaboration})
}
\offprints{dkubas@eso.org}
\institute{ 
   {European Southern Observatory, Casilla 19001, Vitacura 19, Santiago, Chile}
    \and{Astronomisches Rechen-Institut, Zentrum f\"ur~Astronomie, Heidelberg University, M\"{o}nchhofstr.~12--14, 69120 Heidelberg, Germany}
 \and{SUPA, University of St Andrews, School of Physics \& Astronomy, North Haugh, St Andrews, KY16~9SS, United Kingdom}
   \and{University of Notre Dame, Department of Physics, 225 Nieuwland Science Hall Notre Dame, USA} 
    \and{Institut d'Astrophysique de Paris, 98bis Boulevard Arago, 75014 Paris, France}
    \and{Observatoire Midi-Pyr\'en\'ees, UMR 5572, 14, avenue Edouard Belin, 31400 Toulouse, France} 
    \and{Physics department, Faculty of Arts and Sciences, University of Rijeka, 51000 Rijeka, Croatia}
    \and{DSM/DAPNIA, CEA Saclay, 91191 Gif-sur-Yvette cedex, France}
    \and{University of Tasmania, School of Maths and Physics, Private bag 37, GPO Hobart, Tasmania 7001, Australia}
    \and{South African Astronomical Observatory, P.O. Box 9 Observatory 7935, South Africa}
    \and{Perth Observatory, Walnut Road, Bickley, Perth 6076, Australia}
    \and{University of Canterbury, Department of Physics \& Astronomy, Private Bag 4800, Christchurch, New Zealand }
    \and{Niels Bohr Institute, Astronomical Observatory, Juliane Maries Vej 30, DK-2100 Copenhagen, Denmark} 
    \and{Isaac Newton Group, Apartado de Correos 321, E-38700 Santa Cruz de La Palma, Spain}
    \and{Space Telescope Science Institute, 3700 San Martin Drive, Baltimore, MD 21218, USA} 
    \and{Dept. of Physics / Boyden Observatory, University of the Free State, Bloemfontein 9300, South Africa} 
    \and{Technical University of Vienna, Dept. of Computing, Wiedner Hauptstrasse 10, Vienna, Austria} 
    \and{McDonald Observatory, 16120 St Hwy Spur 78, Fort Davis, TX 79734, USA}
\and{Astrophysics Research Institute, Liverpool John Moores University, Twelve Quays House, Egerton Wharf, Birkenhead CH41 1LD, UK}
\and{Department of Astronomy, University of Florida, 211 Bryant Space Science Center, Gainesville, FL 32611-2055, USA}
\and{Lawrence Livermore National Laboratory, IGPP, P.O. Box 808, Livermore, CA 94551, USA}
\and{Astronomy Unit, School of Mathematical Sciences, Queen Mary, University of London, Mile End Road, London E1 4NS, UK}
}

\date{Received ; accepted}
\abstract  
{}
{
We investigate constraints on  additional planets orbiting the 
distant M-dwarf star \system, around which photometric microlensing
data has revealed the existence of the sub-Neptune-mass planet
\planet. We specifically aim to study potential Jovian
companions and compare our findings with predictions from core-accretion and
disc-instability models of planet formation. We also obtain an estimate of the 
detection probability for sub-Neptune mass planets similar to \planet 
using a simplified simulation of a microlensing experiment.}
{ 
We compute the
efficiency of our photometric data for detecting additional planets around \system, 
as a function of the microlensing model parameters and convert it into a function 
of the orbital axis and planet mass by means of an adopted model of the Milky Way.}
{We find that more than $50\%$ of potential planets with a mass in
excess of $1~M_\mathrm{J}$ between 1.1 and 2.3~AU around \system
would have revealed their existence, 
whereas for gas giants above $3~M_\mathrm{J}$ in orbits between 1.5 and 2.2~AU,
the detection efficiency reaches $70\%$;  however,  no such companion was observed.
Our photometric microlensing data therefore do not contradict the existence of gas
giant planets at any separation orbiting OGLE 2005-BLG-390L.
Furthermore we find a detection probability for an \planet-like planet 
of around $2-5\%$.
In agreement with current planet formation theories, this quantitatively 
supports the prediction that sub-Neptune mass planets are common around low-mass stars.}
{}
\maketitle

\section{Introduction} \label{sec:intro}

After more than a decade since the discovery of the first extra-solar planet
orbiting a solar-type star \citep{1995Natur.378..355M} the Extra-solar Planet 
Encyclopedia\footnote{http://exoplanet.eu} lists 249 entries,
including an increasing number (26) of multiple planetary systems. 
{The accessible mass regime of extra-solar planets extends since 2005 below
the $10~\mTer$ regime, with the discoveries
of Gliese~876~d \citep[$\sim7.5~\mTer$,][]{2005AAS...20719103R}
and the pair of planets around Gliese~581 with minimum masses of $5$ and $8~\mTer$
\citep{2007A&A...469L..43U} using the radial-velocity technique, as well as \planet 
detected by microlensing \citep[$\sim5~\mTer$,][]{2006Natur}.}
While recent improvements in radial velocity sensitivity have 
enabled the discovery of Neptune-mass planets in Venus-like orbits
\citep{2006Natur.441..305L,2006A&A...455L..25A}, microlensing is the only method that can
detect such sub-Neptune-mass planets in orbits beyond $1\,$AU.
This sensitivity to sub-Neptune-mass planets at separations of a few AU
is important for testing the core accretion theory of planet formation
because this theory predicts that the dominant planets in any planetary
system should form in the vicinity of the ``snow line\rlap," which is located
at a few AU \citep{2006ApJ...650L.139K,2004ApJ...616..567I,2004ApJ...612L..73L}.
Microlensing results allow this theory to be tested without confronting the
additional uncertainties of planetary migration.

Despite these impressive successes, the observational picture of
the planet abundance and their mass function is far from 
complete. This is partly caused by biases introduced by the detection techniques.
The currently dominant radial-velocity method,
despite ever improving sensitivity and temporal baselines, 
still favors massive planets in close-in orbits around solar type stars within about
$100~\rm{pc}$ from the Sun. The transit method also favors the detection of close-in giant
planets.

The microlensing technique, initially proposed by \cite{1991ApJ...374L..37M} and whose
prospects have been quantified first in more detail by \cite{1992ApJ...396..104G}, has now led to 
four reported detections 
\citep{2004ApJ...606L.155B,2005ApJ...628L.109U,2006Natur,2006ApJ...644L..37G}.
It has proved its capability to provide access to a new window for exoplanets, with 
masses down to an Earth-mass for ground-based searches 
and orbits between $1-10~\rm{AU}$ mainly around 
host stars less massive than the Sun at several kpc distance.   
This is of special interest since most of stars in 
our Galaxy have masses less than $1\,\msun$, so the planets of such stars,
if common, may constitute the majority
of Galactic exoplanetary systems. While the hunt for planets around 
low-mass stars still lags a bit behind the search for planetary companions to solar-type stars, 
the most recent microlensing  
discoveries of sub-Neptune mass (or super-Earth) planets \citep{2006Natur,2006ApJ...644L..37G}
indicate that such planets indeed are common,
probably more common than any other class of exoplanets yet discovered.
The rarity of cool Jovian companions to sub-solar mass stars
(from $1-10~\rm{AU}$) seen in the microlensing data 
\citep{2001ApJ...556L.113A,Gaudi02,2003MNRAS.343.1131T,2004MNRAS.351..967S} on the other
hand was recently complemented by radial velocity searches, deriving an upper limit of
$< 1\%$ for the fraction of close-in (within 1 AU) Jovian M-dwarf planets \citep{2006ApJ...649..436E}. 

These new observational constraints seem to be in line with predictions of
planet formation theories. For example simulations done by \cite{2005ApJ...626.1045I} 
and \cite{2004ApJ...612L..73L} suggest that the formation of Jovian gas giants around M-dwarfs 
is inhibited, while planets less massive than Neptune can easily form. 
This is supported by recent microlensing results as one of the gas-giant planets
discovered by microlensing was recently found to orbit a K-dwarf
\citep{2006ApJ...647L.171B}, 
{while the least massive planets discovered so far (\planet, Gliese~476~b, Gliese~581~c and d) are
all orbiting M-dwarf stars.}
Disc instability formation models as advocated by \cite{2006ApJ...644L..79B} also are
capable of explaining the preference of forming sub-Neptune mass planets rather than giant planets 
around low-mass stars . 

In the study presented here we examine what constraints on a hypothetical  
Jovian planetary companion to the sub-Neptune mass planet detected in the microlensing event 
OGLE 2005-BLG-390 \citep{2006Natur} can be derived from the photometric light curve data. 
This system contains a $3-10~\mTer$ planet in orbit with semi major axis $a=2-5~{\rm AU}$ around an M-dwarf. 
We also estimate the probability of having detected an \planet like planet in an idealized microlensing experiment 
and {discuss in more details} the claims made in \cite{2006Natur} and \cite{2006ApJ...644L..37G} that sub-Neptunes are common companions 
to low-mass stars.

\section{Basics of microlensing}   \label{sec:MCbasis}

\begin{figure*}[ht]
  \begin{center}
       \includegraphics[width=17.cm]{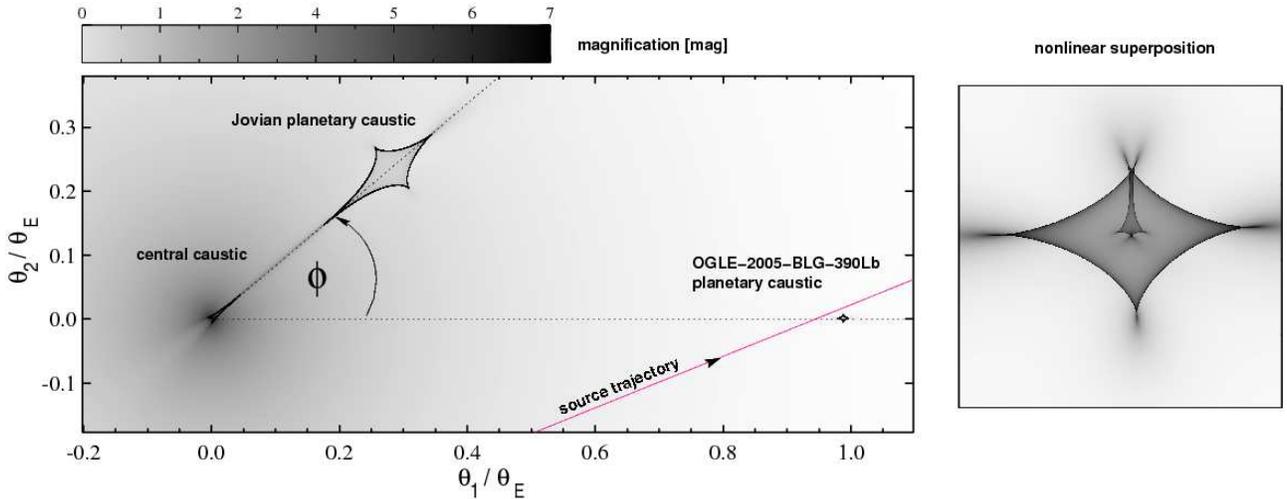}
    \caption{
      {{The left panel of the figure
      displays a magnification map of a triple lens system configuration, consisting of a host star
      at the origin and two planets. The magnification is shown as function of
      the position angle $(\theta_1,\theta_2)$ on the sky. The darker the region the higher the magnification. 
      The caustic contours have been over-plotted (black solid lines) for
      clarity. The small planetary caustic induced by \planet is shown 
       together with the best-fitting source track \cite[ pink solid line]{2006Natur}.
      The larger planetary caustic on the upper left stems from a hypothetical additional Jovian 
      companion, in this particular case with $q=2.0\times 10^{-3}$ and $d=1.2$.
      Note that the actual positions of the planets do not coincide with the position
      of their planetary caustics. They lie on the dotted axes but outside the shown field of view,
      which scale was chosen to maximize the visibility of the caustics. 
      Finally the right panel shows one of the rare configurations where the caustics of the two
      planets are close to the same position merging in a non-linear way, showing the
      limit of the linear superposition approximation.}} 
       }
  \end{center}
 \end{figure*}\label{fig:alpha_demo}

{During a microlensing event, the light arising from a \dan {background star} 
(the {\em source star}) is bent due to the 
gravitational field of an intervening planetary system (the {\em lens}) passing close
to the observer line-of-sight. This results in a characteristic transient magnification of the
source leading to a variation in the received flux, which constitutes the 
microlensing event light curve.}
   
With $M$ denoting the total lens mass and $\DS$, $\DL$ 
the distances of source and lens from the observer respectively, the 
angular Einstein radius \citep{1936Sci....84..506E}
\begin{equation}
   \anger = \sqrt{\frac{4GM}{c^2}\,\left(\frac{\DS-\DL}{\DL\,\DS}\right)}
\end{equation}
defines the natural scale of gravitational microlensing. 
It equals the angular radius of the ring-shaped image 
of the source star which would occur for a perfect alignment between the observer, lens and source. 
With sources typically located  in the Galactic Bulge and a lens 
lying at several kpc from us in the direction of the source, 
its linear scale at the lens distance $\RE=\DL\,\anger$  corresponds
to $\sim 1-10~{\rm AU}$, a range well suited for extrasolar planet hunting.

{The transient brightening of the source (called \emph{magnification}) due to a single lens can 
be fairly easily modeled and is described 
by an impact parameter $\uo$, with $\uo\,\anger$ being the minimal angular distance  
between lens and source at time $\to$, as well as a time-scale $\tE$ which is the duration for the source to
move by $\anger$ on the sky. The total observed flux $F(t)$ is then expressed as the sum of the magnified
source flux $A(t)\times\FS$, with $A(t)$ being the magnification factor, plus the so-called blend flux $\FB$
stemming from unresolved sources within the aperture, 
so that $F(t)=A(t)\,\FS+\FB$. Furthermore, one must take into account a pair of flux parameters ($\FS$, $\FB$) per 
observing telescope.}\\
A lens with a planet is usually modeled by a binary lens with an extreme mass ratio ($q <10^{-2}$).
The most striking characteristic of binary lenses when compared to single lenses is the occurrence of
extended caustics, closed lines defined in the source plane where the magnification becomes
infinite for point sources. However in practice the magnification remains finite since real sources 
have a finite extent. Nevertheless caustics mark regions of large magnification gradients.   
In the case of planetary lensing the area enclosed by these caustics (defining a kind of
cross section) compared to the area
covered by the Einstein ring disc is only a few percent at most, but if the source
trajectory passes sufficiently close to or even traverses a caustic it can imprint a high signal-to-noise 
light curve signature, revealing the presence of the planet. Unfortunately this signal is only
short-lived (ranging from hours to days for Earths to Jupiters) with respect to the complete
event time-scales of weeks to months. This is the big observational challenge for teams like
PLANET/RoboNet\footnote{http://planet.iap.fr\\ $~~~~~~~~$http://www.astro.livjm.ac.uk/RoboNet/}, 
OGLE\footnote{http://www.astrouw.edu.pl/\~{ }ogle}, 
MOA\footnote{http://www.phys.canterbury.ac.nz/moa} and 
MicroFUN\footnote{http://www-astronomy.mps.ohio-state.edu/\~{ }microfun}
to monitor microlensing events with a high sampling rate
and not to miss these planetary `anomalies` in light curves which for most of the time are
indistinguishable from a single-lens event.

The challenge on the modeling side is that the magnification of a binary lens cannot be
expressed in a closed analytical form and its high-dimensional parameter space has an intricate
$\chi^2$-surface on which the problem of parameter optimization is far from trivial.
In addition to the single-lens model parameters the
following ones are required: the planet-to-star mass ratio $q$, the lens separation $d$,
with $d\times\ThE$ being the instantaneous angular distance of the planet from its host star, the
impact angle $\alpha$ between the source trajectory and the binary lens axis.  If the source
is resolved by the lens, the angular source size $\ThS$
and parameters describing the source surface profile also need to be considered.

\section{Detection efficiency of additional planets around \system}\label{sec:def}

{\cite{2000ApJ...528...56G} have presented an algorithm well-suited to compute the ``efficiency''
with which a binary companion to the lens can be revealed in an observed microlensing event. 
This algorithm applies for microlensing events where no clear deviation from a single
point-mass lens can be seen in the light curve, \emph{i.e.} events for which there is no evidence
for a companion to the lens. Efficiency calculations as done in
\cite{2000ApJ...535..176A} or \cite{Gaudi02} indeed were aimed to give limits on
companions to lenses where no planets were found.
Here we extend this method to compute the detection efficiency of multiple planetary systems, more
precisely to put confidence limits on the presence of an hypothetical second planet
in a system where one planet has already been detected.
For similar reasons to the original method, the algorithm holds as long as there is no further
clear signal in the light curve other than the already detected planet.
We first outline the method before applying it specifically to \system.

As a natural extension of the method proposed by \cite{2000ApJ...528...56G}, we use a triple lens
configuration to describe the discovered planet plus and additional companion, for which we want to
compute detection efficiencies (\emph{i.e.} calculating confidence limits on the latter's  existence).
Such a geometrical configuration is shown in the left panel of Fig.~1, where the parent star of the 
system is located at the origin and known \planet is situated on the horizontal axis on the right.
A  hypothetical additional planet with mass ratio $q$ is located at a distance $d$ from
its parent star ($d=1.2$, $q=2\dix{-3}$ and  $\phi=60^{\rm o}$ in  figure) with its 
planet-star axis  subtending an angle $\phi$ with the $x$-axis.
Note that with a planetary signal already detected, the trajectory of the source is known and fixed.
We then compute detection efficiency on the second planet using the algorithm described below:
\begin{enumerate}
\item Compute the $\chi^2_{\rm ref}$ of the model that best fits the already discovered planet,
\item Choose a given triple lens configuration with parameters $(d,q,\phi)$, where
  $\phi\in[0,2\pi]$, and fit for the remaining parameters to compute the $\chi^2$ difference
  $\Delta\chi^2(d,q,\phi)=\chi^2(d,q,\phi)-\chi^2_{\rm ref}$,
\item Repeat step (2) until a dense sampling of $\phi$ is obtained for the probed lens configurations,
\item Repeat from step (3) until all the chosen lens configurations are probed.
\end{enumerate}

The detection efficiency of an additional planet in the system is then given by the fraction of angles $\phi$
that produce a significant deviation in the light curve. By significant
deviation, we mean an excess of $\chi^2$ of the probed lens model by an amount of
$\chi^2_{\rm thresh}$ relative to the single planetary model. More formally the detection
efficiency is given by the formula:
\begin{equation}
  \varepsilon(d,q)=\int\limits_0^{2\pi} H\left(\Delta\chi^2(d,q,\phi) - \chi^2_{\rm thresh} \right)\,\d\phi,
\end{equation}
$H$ being the Heaviside step function. 

Apart from the computationally challenging sheer size of the parameter space of multiple lens systems
another notable difficulty arises when the spatial
extent of the source star has to be considered. Rather than integrating over the source surface, we
make use of magnification maps that are much better designed for our purpose.
A magnification map is obtained by shooting light rays from the observer through the lens and
towards the source \citep{1997MNRAS.284..172W} ; assuming a constant spatial ray density in the 
plane containing the lens, its density in the plane where resides the source then maps the magnification
as a function of its position.
Such maps can easily include the extended source effects: The ray-shot maps just need to be convolved
using the right size and brightness profile of the chosen source 
star \citep[e.g. ][]{2005A&A...435..941K}.
A speed optimized version of the original ray-shooting algorithm is presented in \cite{2002MNRAS.335..159R}. 
Fig.~1 is an example of such a triple lens magnification map.

In general the source size in physical or Einstein units is not known and estimates rely on statistical Galactic models. 
The uncertainty of those parameters directly affects the uncertainties of the computed detection efficiencies \cite[\emph{cf.~}][]{Gaudi02}.
The situation is much more favorable when, as in the case of \planet, extended source effects are detected and strong constraints can be put
on the source properties. Then it is straightforward to convolve a library of maps with the appropiate size and brightness profile of the source
which as a welcomed byproduct reduces the to be explored parameter space.

Nevertheless this approach is still resource intensive since it requires  computing a large number of magnification
maps for the different probed characteristics $(d,q,\phi)$ of the second planet. However, in the case of
\source we can simplify the problem with a fair approximation as follows.
In fact, it has been shown that the planetary caustics in a multiple planetary system are independent 
to first order, as long as the projected planet positions do not overlap \citep{1999A&A...348..311B}.
However when the planetary caustics of the two planets are too close, non-linearities appear
and the superposition principle is not valid anymore. This effect can be seen in the right panel of
Fig.~1, where the two caustics are superposed in a non-linear way. Here the more massive hypothetical planet has been located only 0.05 fractions
of $\RE$ above \planet causing the small caustic to flip its orientation by about 90 degrees and splitting two of its cusps. 
Hence, the superposition principle of the
caustics is valid for most values of the angle $\phi$ and position $d$ of the additional planet,
except when the caustics coincide. 
We make the choice here to compute only binary lens maps and use the superposition principle. By
doing so, we save around two orders of magnitude of computational time to create the map library (each 
magnification map requires few hours of computation with a usual CPU). 

To deal with the rare configurations when
the non-linearities appear,  one choice would be to subtract \dan{ the best-fit binary-lens model from the original \planet light curve 
and then add the residuals to form a new light curve based on the underlying single-lens model. This new "single-lens" light curve 
then serves as a reference to which a series of binary-lens models are fitted according to \cite{2000ApJ...528...56G}.}
However we prefer here to adopt a more conservative choice, by cutting out the data from the region where the signal 
of \planet resides (\emph{i.e} MHJD between $3592.50$ and $3593.11$) 
prior to the computations. While with both options one cannot remove the intrinsic non-linearity at the common
position of the caustics, the uncertainties they introduce are small and in any case of similar 
order of magnitude.

We have computed a $(d,q)$ magnification map library of lens configurations spanning nine different mass
ratios ($q = 5.6\dix{-4} - 10^{-2}$) and covering sixteen lens separations ($d= 0.1 - 3.3$),
convolved with the source brightness profile adopted by \citet{2006Natur}. }
The choice of the grid is motivated by the goal to cover the theoretical range of Jovian type
planets probed in the simulations of \citet{2005ApJ...626.1045I},
and also to ensure a good coverage of the so-called lensing zone, the range of lens separations,
where microlensing is most sensitive to planets. 
For each $(d,q)$ configuration we have computed  $200$ models with 
$\phi_k = 2k\pi/200$, i.e. in total $200 \times 16 \times 9 = 28800$ models. 

At each $(d,q)$-grid point and for each angle $\phi \in[0,2\pi]$ we then compute the least-square
measure $\chi^2(d,q,\phi)$ optimized over the remaining free parameters: minimum impact parameter $\uo$,
time of maximum magnification $\to$, Einstein radius crossing time $\tE$, as well as the source fluxes ${\FS}^{i}$ 
and blending fluxes ${\FB}^{i}$ for each of the 6 different telescope data sets used.
These optimized parameters are `free' in the sense that we allow them to vary within the error bars
of the best-fitting values derived in \cite{2006Natur}. This flexibility minimizes numerical noise in
the calculations without violating the constraints given by the best model. We checked that
the results are consistent with fixing the parameters to their best-fitting values and that
using non-bounded parameters yields unphysical results. 
To carry out the optimization we use a genetics algorithm \citep{1995ApJS..101..309C},
which naturally provides the capability to bound parameters and 
has been shown to explore the intricate parameter space 
of binary lenses more efficiently than classical gradient optimization techniques \citep{2005PhDTanguero,2005PhDCassan}. 

A given choice of parameters $(d,q,\phi)$ is considered to produce
a significant deviation if $\Delta\chi^2 = \chi^2(d,q,\phi) - \chi^2_{\rm ref}$ 
exceeds a threshold value $\chi^2_{\rm thresh}$. We set  $\chi^2_{\rm thresh}=60$, which we find robust enough to
avoid false detections arising from statistical fluctuations or unrecognized low-level systematics and to
be consistent with earlier detection efficiency studies \citep{Gaudi02,2004ApJ...616.1204Y}.

\Fig{fig:result_raw} shows contours of the detection efficiency $\varepsilon(d,q)$ as a function of the 
dimensionless model parameters $d$ and $q$, where $d\times\ThE$
is the angular separation of the planet from its host star while
$q$ is the planet-to-star mass ratio. {Calculations using an algorithm based on a
method described in \cite{2000ApJ...533..378R} were also used to compute detection efficiencies, 
with comparable results.

\begin{figure}[!h]
  \begin{center}
  \includegraphics[width=9.1cm]{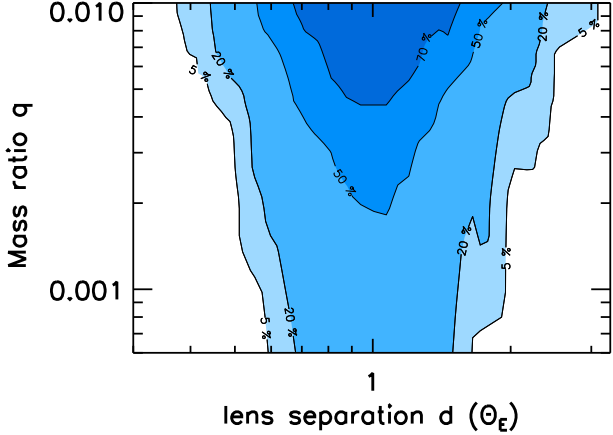}
  \caption{
   Detection efficiency $\varepsilon(d,q)$ as a function of angular planet-to-star separation
   $d\times\ThE$ and planet-to-star mass ratio $q$. These dimensionless efficiencies present
   the raw outcome of our computation prior to the convolution given in Appendix A where we derive
   the physical detection efficiency values as functions of planet mass and orbital radius. 
 }
\label{fig:result_raw}
\end{center}
\end{figure}

The obtained detection efficiency constraints are admittedly rather weak. However this is
not a flaw of the data or analysis but mainly due to the intrinsic nature of a low magnification event.
In fact, \cite{2000ApJ...528...56G} have shown that the detection efficiency of a microlensing event
is strongly dependent on the minimum parameter $\uo$ of the underlying single lens model:
the smaller $\uo$, the higher the peak
magnification and the higher the detection efficiency. This can be understood as follows. 

As depicted in Fig.~1, in a planetary binary lens scenario
one has to differentiate between the central caustic, always located
close to the primary lens (star), and the planetary caustics, the location of which strongly depends on the 
projected star-planet separation $d$. While as stated earlier the planetary caustics are basically mutually
independent from each other, the central caustic is affected by any companion to the primary lens and
thus also especially sensitive to multiple planet systems. To probe the central caustic sufficiently small
impact parameters $\uo$, respectively high magnifications, are required.\cite{1998ApJ...500...37GS}
showed that for  $\uo \rightarrow 0$ the planet detection efficiency goes to one, 
unless finite-source effects prevent the detection of the least massive planets around the highest peaks 
for larger sources.
The drawback of high magnification central caustic events however is that they 
are much harder to model and sometimes plagued by \dan {a close-wide binary ambiguity} 
\citep{Dominik1999, 2005A&A...435..941K}. In the \event event the impact parameter is too 
large to explore the central caustic, which explains the calculated restricted sensitivity to 
additional planets.}

\section{Limits on Jovian companions to \system}   \label{SEC:limits}

While $\Epsdq$ is the most straightforward constraint we can get from modeling, one would like
to infer statements on the underlying physical parameters, the planet's orbital axis $a$ and its mass $m$.
The planet mass and the mass ratio are linked by $\mpla = q\times\mstar$, with $\mstar$ being the host star mass. 
The physical (instantaneous) lens separation projection $r$ is given by  $r = d\times\DL\times\ThE$.
Adopting the median values determined by \citet{2006Natur}, 
namely $\DS \simeq 8.0\,\mathrm{kpc}$, $\DL \simeq 6.6\,\mathrm{kpc}$ 
and $\mstar \simeq 0.22\,\msun$, we have:
\begin{equation}
  \begin{array}{l}
    \mpla = q\times\mstar \simeq q\times0.22\,\msun \\
    r = d\times\DL\ThE \simeq d\times1.35\,\mathrm{AU}.
  \end{array}
\end{equation} 
However, without a proper measurement of both the source size and the parallax in Einstein radii \cite[eg.][]{Ghosh2004,Jiang2004},
the mass $M$ of the host star OGLE 2005-BLG-390L and the Einstein radius are distributed in
probability over a finite range, where both distributions are correlated, and moreover, the orbital axis 
follows from $r$ by a stochastic orbital de-projection. 
Therefore, one needs to take these distributions into account for expressing the detection efficiency as function of
$(a,m)$. Assuming circular orbits and the stellar mass function from \cite{Chabrier:massfunc}, 
we use the implementation of a Galactic model by \cite{Do:Estimate2} in order to derive detection efficiency values
$\varepsilon(a,m)$. The details of this procedure are given in Appendix A.

\begin{figure}[!h]
\begin{center}
   \includegraphics[width=9.1cm]{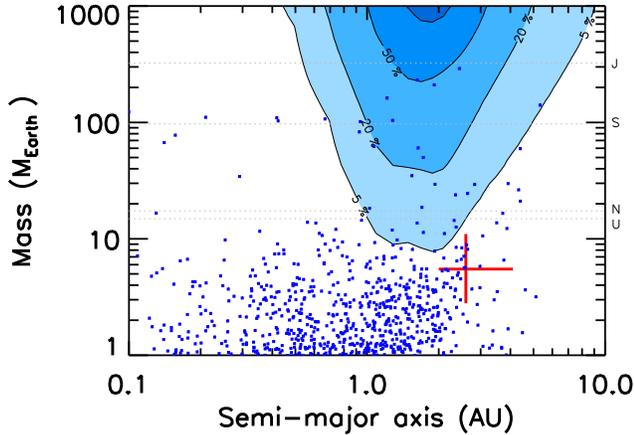}   
   \caption{
    Detection efficiency $\EpsphysM$ for additional
    planets  orbiting \system as function of orbital
    separation $a$ and planet mass $m$, where contours at $5, 20, 50$ and $70~\%$ are shown. 
    The cross marks the median values for the properties of OGLE 2005-BLG-390Lb along with $68.3\%$
    confidence intervals and the dashed horizontal 
    lines mark the masses of Jupiter (J), Saturn (S), Neptune (N) and Uranus (U) for comparison. 
    The blue dots represent the predicted final distribution
    of a seed of $20000$ planetary cores around an M-dwarf of $0.2\msun$ resulting from
    a core-accretion model assuming inefficient migration (taken from Fig.~9b of \idalin).}
  \label{fig:idalin_and_us}
\end{center}
\end{figure}

The resulting detection efficiency diagram is presented in \Fig{fig:idalin_and_us}, which shows a contour plot of
$\varepsilon(d,q)$ together with \planet (red error bar cross).
It basically tells us about confidence limits on additional companions to \planet: values of $\varepsilon(d,q)$ close
to unity rule out the possibility of such an additional planet at $(a,m)$, while
values close to zero mean that no conclusion can be drawn about the multiplicity of \system.

The strongest constraints from the diagram are placed in the
Jovian-mass regime, where $\varepsilon(a,m) > 50~\%$.
We find that planets more massive than $1~M_\mathrm{J}$ in the
orbital range $1.1-2.3$~AU have a detection efficiency in excess of $50~\%$, while
planets above $3~\MJ$ in orbits between $1.5$ and $2.2$~AU would have revealed 
their existence with a probability of more than $70~\%$.
Unfortunately, our data cannot tell us about possible Pegasid planets
(giant planets orbiting at fractions of $1~$AU), which could exist in \system. 

On the same \Fig{fig:idalin_and_us}, we have plotted the
results of Monte-Carlo simulations by \idalin, producing the final
evolution stage of a seed of $20000$ planetary embryos, uniformly
distributed in $\log a$ (from $0.1$ to $100$~AU), around a host star
with $M = 0.2~M_{\sun}$, similar to OGLE~2005--BLG--390L.
In principle, such simulations are valid when considering a single
planet, whereas we consider the case of two planets. 
We therefore choose among the available models of \idalin one for which
the planet migration process is very inefficient and so practically
assume that the planets were formed quasi-\emph{in situ}. This
allows us to use the efficiency diagram to compare our observation to this planet
formation model.
In the chosen planet formation scenario -- model from Fig.~9b of
\idalin -- migration is strongly suppressed, the cores have more time for accretion,
more gas giants can form and planets  (blue points on \Fig{fig:idalin_and_us}) stay close to their orbital birth places. 
Two main predictions can be read out from their simulation.
Firstly, sub-Neptune-M-dwarf planets should vastly outnumber gas-giant-M-dwarf ones. In the case of 
an $0.2~M_{\sun}$ M-dwarf the formation of a sub-Neptune planet between 1 and 14 $\mTer$ in
the orbital range of 1-10 AU is about $\sim 1200$ times more likely than forming Jovian 
planets with $0.5-1~M_{\rm J}$ in the same orbital range.
Secondly if giant planets form they are unlikely to become more massive than $~\sim 1~M_{\rm J}$.

In the framework of disc instability planet formation theory, the 
preference for sub-Neptune planets can also be explained by photo evaporation of the gas envelopes
of giant protoplanets, as recently pointed out
by \cite{2006ApJ...644L..79B}. However in this case  there are no obvious
reasons for postulating a gap in the planetary mass function between Jupiter and Neptune masses.

We note that also around the other known low-mass planet found by microlensing \planetgould,
which host-star is most likely a K-dwarf, there were no traces found in the data of additional Jovian 
companions in the orbital range between about 0.5 and 14 AU \citep{2006ApJ...644L..37G}.  

The current sample of known planets around M- and K-dwarfs is still too small to
distinguish between the alternative scenarios of planet formation and
migration. However, just a few more detections are likely to add valuable
further information.

\section{On the detection of \planet-like planets}  \label{SEC:detprob}

\cite{2006Natur} and \cite{{2006ApJ...644L..37G}} have stated
that their respective detections of the low-mass planets 
\planet and OGLE~2005-BLG-169Lb imply a large abundance of their
siblings. The sensitivity of a microlensing light curve to the presence of
planets depends strongly on, besides data sampling and quality, 
the event magnification and to some extent on
the event time-scale and source size as well. 
Increasing the angular size of the source has, for instance, the dual
effect of increasing the duration of the planetary signal while, at the same time, decreasing its amplitude.
The latter effect ultimately restricts the capabilities for 
detection of planets to $\sim 1~\mTer$ for giant sources and to $\sim 0.1~\mTer$ for 
turnoff stars in the Galactic bulge \citep{1996ApJ...472..660B}.  

An accurate determination
of planet abundances therefore requires an analysis of the detection efficiency
of a representative sample of the whole experimental data set
and a comparison with the actual detections. For the PLANET
campaign, this will be worked out in an upcoming study 
\citep{2006longCassan}. Here, we run simulations to derive estimates of the 
probability for detection of sub-Neptune or Jovian planets based on simplified assumptions.

We choose to pick impact parameters $\uo$ from a uniform distribution, which is close
to what is actually realized in nature ; however, only those events with a magnification above a characteristic threshold will be observable.
Neglecting blending, we sample $\uo \in [0,1]$ in 100 equally spaced steps, corresponding to magnifications
$A>1.34$. While this means that we practically cover the whole
range of magnifications in the alerted and followed-up microlensing events, we neglect 
the preference for higher magnifications in the follow-up campaigns.

In \Sec{sec:def} we made use of a $\Delta\chi^2 $-based criterion for ruling out
additional planets in the \system system. While in principle such a criterion works for detections as well
we note that in practice one often finds that even if this criterion is fulfilled, a lack in coverage and/or 
precision makes it impossible to characterize the planetary model unambiguously. 
Unlike a rejection a convincing detection has to meet stronger criteria
(as, for instance, not showing systematic trends and having a sufficient number of 
data points \citep{2000MNRAS.319.1011V}), 
which are not reflected in a criterion solely based on $\chi^2$. In other words we want to avoid cases in which 
planets are detectable but not characterizable, respectively cannot be distinguished from non-planetary solutions.  

To minimize the effect of this caveat on our simulation we therefore adopt a more demanding 
criterion to calculate the efficiency to discover \planet-like planets.
For a ``discovery'', we demand that the planetary signal amplitude exceeds 
$2\%$ for at least {15 measurements, assuming an hourly sampling and a photometric accuracy of  $1\%$.
Neglecting nonwhite noise this criterion then translates into $\Delta\chi^2 =60$, i.e. is consistent with 
the detection threshold chosen in  \Sec{sec:def}. 
This criterion ensures  that the planetary light curve signature is also well sampled, 
which is essential for a proper characterization of the lens system.
For comparison we also compute the efficiencies for the following set of criteria:
The signal amplitude should exceed $2\%$ for at least  10 and 25 measurements.
}

Since the detection efficiency reaches a maximum in the 
{so-called ``lensing zone'',
corresponding to a star-planet separation range of}
$0.62 \leq d \leq 1.62$, we can restrict our sampling to the
time range $[\to-\tE,\to+\tE]$, as most microlensing 
campaigns do, which ensures being sensitive to planets at these
separations. To avoid border effects, we however draw the simulated light curves from
magnification maps spanning $\to \pm 1.2\,\tE$.
For the source star, we adopt an angular size
$\ThS = 9.6 \dix{-3}\ThE$ (which corresponds to $\ThS = 5.25~\mu\mbox{as}$ or a
physical radius $R_\star \sim 10~R_\odot$ in the case of \planet).

As in Sect.~3, we use the ray-shooting technique for calculating magnification maps. 
While we adopt fixed mass ratios of $q=7.6 \dix{-5}$, which is the value for \planet, 
and $q=4.3 \dix{-3}$, which corresponds to $M = 1~M_\mathrm{J}$ if the mass of \planet
is $5.5~M_\oplus$, we use a grid for the lens separation $d = 0.1 - 5.0$. Again as in \Sec{sec:def}, we compute
the fraction of detections made over this grid assuming that every lens
has such a planet by averaging over the impact angle and impact parameter.

\begin{figure}[!h]
  \begin{center}
       \includegraphics[width=9.1cm]{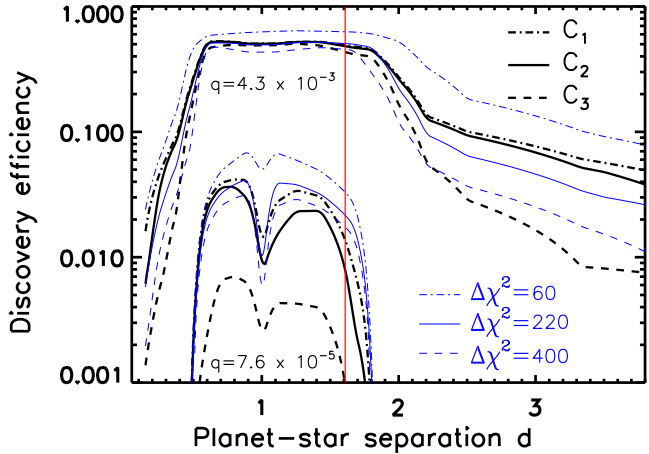}
   \caption{{}
    { Efficiency of detecting a super-Earth planet with the mass ratio of \planet 
    ($q=7.6 \dix{-5}$, lower set of curves), 
    and for  a giant planet with a Jupiter/Sun mass ratio ($q=4.3\dix{-3}$, upper set of curves) 
    as a function of the projected planet-star separation $d$. 
    {The bold black  curves have been computed }
    assuming {an hourly sampled light curve with $1\%$ photometric accuracy}, and demanding 
    that the planetary 
    {signature amplitude exceeds $2\%$ for at least 10 measurements } 
    (dotted line, indexed as  $C_1$), 15 and 25 measurements (solid line, $C_2$ and dashed line, $C_3$).
    For comparison efficiencies for the same sampling rate and accuracy but 
    using a pure $\Delta\chi^2 $ threshold are plotted as well, with $\Delta\chi^2 = 60, 220, 400$
    (blue thin lines).
    The considered impact parameters are drawn uniformly from $\uo \in [0,1]$.
    { Based on the parameters found for the lens scenario of \event, }
    a time scale of $\tE=11.0~ $days was adopted, { as well as a
    source size of $\sim 10~\rsun$.}
    The vertical line marks the {actual} separation of \planet. }}
   \label{fig:detprob390}
\end{center}
\end{figure}

In \Fig{fig:detprob390} the derived efficiencies are shown as
a function of the projected separation $d$, for the 
selected mass ratios 
{and time scale, $\tE=11.0$ as measured for \event.  For  criterion  $C_2$ (more than 15 deviating points) 
at the separation parameter for \planet, $d = 1.61$, the detection
efficiency for an \planet-like mass ratio of $q = 7.6 \dix{-5}$
is about $1\%$, whereas it becomes about $50\%$ for $q = 4.3 \dix{-3}$, resembling a Jupiter-mass planet
around the same host star, i.e.  the sensitivity to Jupiters in this case is, as was reported 
by \citet{2006Natur}, about 50 times higher than it is to sub-Neptunes. 
Unsurprisingly the sensitivity to Jovians only shows a weak dependance on the applied detection criteria, since
the majority of the associated anomalies are strong enough to be detected even by our most stringent detection demands.
The sensitivity for  \planet-like mass ratio on the other hand is much more affected by the choice of the detection threshold
and, when integrated over the lensing zone, ranges from $\sim0.5-3~\%$  for criteria $C_3$ to $C_1$. 
For comparison we have also computed and plotted the detection efficiency using a pure $\Delta\chi^2$ threshold
with $\Delta\chi^2=60,220, 400$, assuming the same light curve sampling rate (hourly) and photometric accuracy ($1~\%$).
We note that whereas the efficiencies for Jovian planets again remain rather unaffected at levels of $\sim 50-60 \%$,
the detection efficiencies drops to $2-5~\%$ for \planet-like companions. 
In fact, a criterion assuming $\Delta\chi^2=60$ is much less demanding that the $C_2$ criterion,
since the latter rejects a lot of anomalies which are too short lived to pass the minimal sampling requirement of 15 points.

Averaged over the lensing zone one finds sensitivity ratios between Jupiters and sub-Neptunes
of up to 25 for \event-like time scales of  $\tE=11.0$ days. For time scales of around $\tE=22.4$ days, which is the median
time scale of events followed up by planet,  the simulations yield that Jovians are about 10 times easier 
to detect that sub-Neptunes.
}

However, there are some notable differences between the 
assumptions for our simulation and the real campaigns.
First,  even in dense telescope networks, weather and technical losses
are unavoidable and a continous data coverage can not always be achieved.      
Secondly the peak magnification of \planet ($A_{\rm o} \sim 3$, corresponding to $\uo = 0.36$) 
is one of the lowest among the events monitored by PLANET/RoboNet. Follow-up teams preferentially 
monitor events with larger magnifications, thereby preferring smaller $\uo$.
As stated at the end of Sec.~3, the detection efficiency usually increases with the magnification and 
therefore the  calculated efficiency for $\uo$ uniformly chosen in $[0,1]$  underestimates the chance to detect
less-massive planets in high magnification events. 
{Moreover, campaigns like PLANET/RoboNet intensify the sampling over the peak region, 
which significantly enhances the chance 
to detect and characterize super-Earths as compared to a standard sampling scheme.
On the other hand,  
our simulations overestimate the chance to detect \planet-like in low magnification events, 
due to the increased anomaly durations when a giant source sweeps over a planetary caustic, whereas a main sequence source star
would on the contrary produce a  signal with higher amplitude but shorter duration.}

Despite its simplicity, our simulation however shows robustly that
in any case \planet-like planets have a modest but not negligible
probability for discovery, while Jovian-type planets should be found
much more easily. With \planet already being the third reported
detection of a planet by microlensing \citep{2006Natur} after two Jovian planets, we therefore have the 
first observational evidence for the suggestion that sub-Neptunes are common around
M-dwarfs and much more frequent than Jovian companions. The detection
of another super-Earth by \cite{2006ApJ...644L..37G} gives further support to this finding.

\section{Summary and conclusions}

We have re-examined the photometric data on \planet to look for traces of additional Jovian companions, finding
that additional planets more massive than $1~M_{\rm J}$ and in orbits of $\sim 1.1-2.3$~AU would have caused
a detectable signal in more than $50\%$ of the cases, which however was not observed.
Planets with  masses of $3~M_{\rm J}$ and above between $\sim~1.5-2.2$~AU would have revealed 
themselves in the data with a probability of $70\%$.

Planet formation models based on sequential accretion processes by
\idalin predict that the creation of gas giants is strongly
suppressed around M-dwarfs for practically the whole range
of their model parameters. 
\dan {In agreement with such theoretical predictions we do not find in our data
any indication of an additional companion to OGLE~2005--BLG--390Lb in the Jovian mass regime, 
however we are not able to definitely exclude this possibility.}

Assuming the natural unfiltered uniform distribution of lens-source impact
parameters above an un-blended magnification threshold of $A = 1.34$ and
a simple sampling pattern, we find a detection probability for an
OGLE 2005-BLG-390Lb-like planet in events involving giant source stars
of $1-3\%$ at its angular separation of 1.61 times the angular Einstein
radius $\ThE$, whereas the average over the lensing zone, \emph{i.e.} 
separations between $0.62$ and $1.62\,\ThE$, becomes $2-5\%$. 
With detection probabilities of a few percent, our discovery of \planet provides the first observational evidence in support of the
prediction by current planet formation theories that sub-Neptune mass planets are common.

\begin{acknowledgements}
We are very thankful to  S.~Ida and D.N.C.~Lin for providing their simulation results and permission to
incorporate them in Fig.~\ref{fig:idalin_and_us}. We express our deepest gratitude to the OGLE and
MOA collaborations for continuing to provide us with alerts on ongoing microlensing events from which we
choose our targets. Part of the work was supported by an ESO DGDF grant for D.K. and a University
of Tasmania IRIG grant for D.K. and A.C. We especially thank the University of Tasmania 
for granting us access to their TPAC supercomputer facilities where part of the calculations were carried out. 
This work was also supported by the French HOLMES ANR grant. 
KHC's work as performed under the auspices of the US DOE, LLNL in part
under Contract W-7405-Eng-48 and in part under Contract DE-AC52-07NA27344.
We also would like to thank the anonymous referee for his constructive comments.
Furthermore D.K. and A.C. would like to thank C.~Mordasini for
inspiring discussions on planet formation theory. 
\end{acknowledgements}

\bibliographystyle{aa}
\bibliography{def390}

\begin{appendix}
\section{Conversion to physical lens parameters}

The parameters that can be extracted from the observed light curve are
not sufficient for directly determining the properties of the planet and its host star, such as their masses,
the orbital radius and period, as
well as their distance. However, as discussed by
\citet{Do:Estimate2}, one can derive probability densities
by means of Bayes' theorem under the assumption of a mass spectrum for the
lens stars, the lens and source distance, following the spatial mass 
density of stars in the Milky Way, and their velocity distribution. 
The planet itself is characterized by the planet-to-star mass
ratio $q = m/M$ and the separation parameter $d$, where $d\,\theta_\mathrm{E}$ is the 
instantaneous angular separation between the planet and its
host star. All further information relies on the time-scales
$t_\mathrm{E} = \theta_\mathrm{E}/\mu$ and 
$t_\star = \theta_\star/\mu$, during which the source moves
by the angular Einstein radius $\theta_\mathrm{E}$ or its own
angular radius $\theta_\star$ relative to the lens on the sky, where
$\mu$ denotes its relative proper motion. With $\theta_\star$ determined from 
the source magnitude and color, one obtains $\mu$ and $\theta_\mathrm{E}$.

With $D_\mathrm{L}$ and $D_\mathrm{S}$ denoting the lens or source
distance, respectively,
let us define fractional distances $x = D_\mathrm{L}/D_\mathrm{S}$ and
$y = D_\mathrm{S}/R_\mathrm{GC}$, where $R_\mathrm{GC} = (7.62 \pm 0.32)~\mbox{kpc}$ is the distance 
to the Galactic center \citep{2005ApJ...628..246E}.
If one ignores selection effects for the source stars related to their
luminosity function as well as extinction, the probability of finding
these between $D_\mathrm{S}$ and $D_\mathrm{S} + \mathrm{d}D_\mathrm{S}$
becomes proportional to $D_\mathrm{S}^2\,\rho_\mathrm{S}(D_\mathrm{S})$,
where $\rho_\mathrm{S}(D_\mathrm{S})$ denotes the volume mass density of
the source stars, while $\rho_\mathrm{L}(D_\mathrm{L})$ denotes the volume mass density of the lenses.

We further assume that the lens star mass spectrum $\Phi_{M/M_{\sun}}
(M/M_{\sun})$ stretches from $M_\mathrm{min}$ and $M_\mathrm{max}$ and
does not depend on the lens distances, while we consider the 
distribution of the effective lens velocity $v = D_\mathrm{L}\,\mu$
to depend on both the lens and source distance. With a characteristic
velocity $v_\mathrm{c}$, the dimensionless velocity parameter
$\zeta = v/v_\mathrm{c}$ is distributed as $\Phi_\zeta(\zeta,x,y)
= v_\mathrm{c}\,\Phi_v(v_\mathrm{c}\,\zeta,x,y)$.

With 
\begin{eqnarray}
\theta_{\mathrm{E},{\sun}} & = &
2\,\sqrt{\frac{G\,M_{\sun}}{c^2\,R_\mathrm{GC}}}
= 1030 \left(\frac{R_\mathrm{GC}}{7.62~\mbox{kpc}}\right)^{-1/2}
\mbox{{$\mu$}as} \\
\eta_{\theta_\mathrm{E}} & = & \frac{\theta_\mathrm{E}}{\theta_{\mathrm{E},\sun}} = 9.68 \times 10^{-4} \left(\frac{\theta_\mathrm{E}}{1~\mbox{$\mu$as}}\right)\,\left(\frac{R_\mathrm{GC}}{7.62~\mbox{kpc}}\right)^{1/2} 
\\
\eta_\mu & = & \frac{\mu\,R_\mathrm{GC}}{2\,v_\mathrm{c}}
= 0.0660\,\left(\frac{\mu}{1~\mbox{$\mu$as}\,\mbox{d}^{-1}}\right)\,
\left(\frac{R_\mathrm{GC}}{7.62~\mbox{kpc}}\right)\,,
\end{eqnarray}
the measured values of $\theta_\mathrm{E}$ and $\mu$ determine
the detection efficiency as function of the projected separation
$ r = d\,D_\mathrm{L}\,\theta_\mathrm{E}$ and the planet mass
$m$ as
\begin{eqnarray}
& & \hspace*{-1em} \varepsilon_{ r,m}( r,m) 
=  \int\limits_{\min(m/M_{\sun},M_\mathrm{min}/M_{\sun})}^{M_\mathrm{max}/M_{\sun}} \int\limits_0^{\infty} \; \times \nonumber \\
& & \quad \times \;
\varepsilon_{d,q}\left[\frac{ r}{y\,R_\mathrm{GC}\,\eta_{\theta_\mathrm{E}}\,\theta_{\mathrm{E},{\sun}}}
\left(\frac{y\,\eta_{\theta_\mathrm{E}}^2}{M/M_{\sun}}+1\right),
\frac{m/M_{\sun}}{M/M_{\sun}}\right] \; \times \nonumber \\
& & \quad \times\; p_{M/M_{\sun}}(M/M_{\sun},y; \eta_{\theta_\mathrm{E}},
\eta_\mu)\,\mathrm{d}y\,\mathrm{d}(M/M_{\sun})\,,
\end{eqnarray}
where $p_{M/M_{\sun}}$ denotes the probability density of the mass of the primary, which reads
\begin{eqnarray}
& & \hspace*{-1em} p_{M/M_{\sun}}(M/M_{\sun},y;\eta_{\theta_\mathrm{E}},\eta_\mu) 
= \frac{R_\mathrm{GC}^2}{N_0\,M_{\sun}^2}\,f_y(y)\,y^{9/2}\,\eta_\mu\,\eta_{\theta_\mathrm{E}}^2\;\times \nonumber \\
& & \quad \times \;
\frac{\left(M/M_{\sun}\right)^3}{\left(y\,\eta_{\theta_\mathrm{E}}^2 + M/M_{\sun}\right)^5}\,\Phi_{M/M_{\sun}}(M/M_{\sun}) \; \times \nonumber \\
& & \quad \times \;
\Phi_\zeta\left(\frac{2\,y\,\eta_\mu\,(M/M_{\sun})}{y\,\eta_{\theta_\mathrm{E}}^2 
+ M/M_{\sun}},\frac{M/M_{\sun}}{y\,\eta_{\theta_\mathrm{E}}^2+M/M_{\sun}},y\right)\; \times \nonumber \\
& & \quad \times \;
\rho_\mathrm{L}\left(\frac{M/M_{\sun}}{y\,\eta_{\theta_\mathrm{E}}^2 + M/M_{\sun}}\right)\,
\rho_\mathrm{S}(y\,R_\mathrm{GC})\,,
\end{eqnarray}
where $N_0 = \int\!\!\int p_{M/M_{\sun}}(M/M_{\sun},y;\eta_{\theta_\mathrm{E}},\eta_\mu)\,\mathrm{d}y\,\mathrm{d}(M/M_{\sun})$.

The function $f_y(y)$ denotes a prior for the source distance, for which 
we adopted
\begin{equation}
f_y(y) = \frac{1}{\sqrt{2\,\pi}\,\sigma_y}\,
\exp\left\{-y^2\,\frac{[\ln y-\ln y^{(0)}]^2}{2\,\sigma_y^2}\right\}
\end{equation}
so that $\ln y$ follows a normal distribution
around $\ln y^{(0)}$ with standard deviation $\sigma_{\ln y} = \sigma_y/y$.

With the assumption of circular orbits, de-projection assuming random 
orientation and phase, finally yields the detection efficiency as
function of the orbital radius $a$ and the planet mass $m$ as
\begin{equation}
\varepsilon_{a,m}(a,m) = \int \limits_0^1 \varepsilon_{\hat r,m}(\sqrt{1-{\hat w}^2}\,a,m)\,\mathrm{d}{\hat w}\,.
\end{equation}

For the mass and velocity distribution of the Milky Way, we follow
\citet{Do:Estimate2} in adopting a double-exponential disk and a barred bulge as well as mass spectra 
for disk and bulge lenses discussed by
\citet{Chabrier:massfunc}. 

For OGLE-2005-BLG-390, we inferred $t_\mathrm{E} = 11.0~\mbox{d}$ and
$t_\star = 0.28~\mbox{d}$, so that with $\theta_\star = 5.25~\mbox{$\mu$as}$, 
one finds $\mu = 18.6~\mbox{$\mu$as}\,\mbox{d}^{-1}$ and
$\theta_\mathrm{E} = 205~\mbox{$\mu$as}$. 
Typing of the source by \citet{2006Natur},
yielded $D_\mathrm{S}^{(0)} = 8.0~\mbox{kpc}$ and $\sigma_{D_\mathrm{S}} = 2.0~\mbox{kpc}$.

\end{appendix}

\end{document}